\documentclass[aps,prl,reprint,twocolumn,notitlepage,superscriptaddress,nofootinbib]{revtex4-1}
\usepackage{amssymb,bm,bbm}
\usepackage{amsmath,eucal}%,cancel}
\usepackage[usenames,dvipsnames]{color}
\usepackage{pdfpages}
\makeatletter
\AtBeginDocument{\let\LS@rot\@undefined}
\makeatother

\usepackage{hyperref}
\hypersetup{
    colorlinks=true,
    linkcolor=blue,
    filecolor=magenta,      
    urlcolor=cyan,
}

\usepackage[table, svgnames, dvipsnames]{xcolor}
\usepackage{multirow,graphicx,booktabs,colortbl}

\definecolor{lb}{rgb}{.804,.851,.922}

\begin{document}

\title{Chemotaxis emerges as the optimal solution to cooperative search games}

\author{Alberto Pezzotta}
\affiliation{International School for Advanced Studies (SISSA)}
\author{Matteo Adorisio}
\affiliation{International School for Advanced Studies (SISSA)}
\author{Antonio Celani} 
\affiliation{The Abdus Salam International Centre for Theoretical Physics (ICTP)}


\begin{abstract}
Cooperative search games are collective tasks where all agents share the same goal of reaching a target in the shortest time while limiting energy expenditure and avoiding collisions. Here we show that the equations that characterize the optimal strategy are identical to a long-known phenomenological model of chemotaxis, the directed motion of microorganisms guided by chemical cues. Within this analogy, the substance to which searchers respond acts as the memory over which agents share information about the environment. The actions of writing, erasing and forgetting are equivalent to production, consumption and degradation of chemoattractant. The rates at which these biochemical processes take place are tightly related to the parameters that characterize the decision-making problem, such as learning rate, costs for time, control, collisions and their trade-offs, as well as the attitude of agents toward risk. We establish a dictionary that maps notions from decision-making theory to biophysical observables in chemotaxis, and vice versa. Our results offer a fundamental explanation of why search algorithms that mimic microbial chemotaxis can be very effective and suggest how to optimize their performance. 
\end{abstract}

\maketitle

\paragraph*{Introduction.}
Individuals in a group often have to face complex situations which require concerted actions \cite{bonabeau-97, bonabeau-99, garnier-07}. Among the various collective intelligence problems, here we focus our attention on cooperative navigation tasks, where all agents share the common goal of locating a target and reaching it in the most efficient way. For instance, a crowd may need to quickly escape from an enclosed space while averting stampedes. Similarly, birds in a flock or fish in a school try to reduce exposure to predators and avoid harmful collisions. In addition, individuals are also confronted with the limits posed by the energetic costs of locomotion. The very same kind of objectives and challenges lie at the heart of multi-agent autonomous robotics\cite{panait-05, viragh-14, quadrotor-15, hsieh-18}. 

Intelligent agents should aim at acting optimally in these contexts. That is, they should cooperate in order to minimize some cost function that compounds the many objectives at play: short time for completing the task, small energy spent in the process, and reduced damage by collisions. What is the optimal strategy? How universal is it across environments and agents? How is information shared by agents? How is it translated into actions?  Can the optimal behavior be reliably and quickly learned by agents facing unknown environments? Is the optimal strategy actually employed by living organisms? 

In this paper we answer these questions by formulating the cooperative search game in terms of stochastic optimal control. We first discuss how optimal solutions can be mapped into quantum states of an interacting many-body system. Unfortunately, the exact solution of this quantum problem is  very difficult even in simple geometries. However, 
in the limit of very large collectives, a mean-field theory  
yields very simple and well-known effective equations. 

Indeed, %somewhat surprisingly, 
the mean-field equations for optimal cooperative search turn out to be identical to a long-known phenomenological model of chemotaxis, the celebrated Patlak--Keller--Segel model \cite{patlak-53,ks-71} with Weber--Fechner logarithmic response (see e.g. \cite{adler-17} for a general discussion about fold-change detection). The chemical attractant can therefore be interpreted as the medium that agents use to share information about the location of the target and the density of individuals in the group. 
The biophysical processes by which the concentration is altered, namely production, consumption and degradation, correspond to the actions of writing information on the memory, erasing and forgetting, respectively. 
We show that there is a dictionary that maps concepts from decision-making theory -- strategies, desirability, costs for control and for collisions, cost per time elapsed, attitude toward risk -- into precise physico-chemical and biological correlates -- concentration levels, diffusion coefficients, degradation and consumption rates, chemotactic coefficients (see Table~\ref{tab:corr} for the detailed analogy).

\begin{table*}[ht]
\centering
\begin{tabular}{|>{\centering}m{.07\textwidth}>{\raggedright}m{.22\textwidth}|>{\centering}m{.20\textwidth}m{.38\textwidth}|}
	\hline
	\multicolumn{2}{|c|}{\textbf{Decision making}} 
		& \multicolumn{2}{c|}{\textbf{Chemotaxis}} \\
	\hline
%	\rowcolor{lb}
		$\zeta$ & Desirability & $s$	& Chemoattractant concentration \\ 
	\hline
	$D$ & Uncontrolled dynamics & $D$ & Random motility\\ 
	\hline
%	\rowcolor{lb}
		$u^*$     & Optimal control 		& $\chi \nabla \log s$ & Chemotactic drift with logarithmic sensing\\
	\hline
	$\gamma$ & Weight for the cost of control & \multirow{2}{*}{$\chi={2D}/{(1-2D\alpha\gamma)}$} & \multirow{2}{*}{Chemotactic coefficient} \\
	$\alpha$ & Risk sensitivity &  &  \\ 
	\hline
	%\rowcolor{lb}
		$q$     & Time cost rate  &  $D_s=D/\epsilon$ & Diffusion coefficient of chemoattractant \\
	%\rowcolor{lb}
	    $g$     & Collision cost rate &   $k=q (1-2D\alpha\gamma)/2D\gamma\epsilon$ &  Degradation rate of chemoattractant \\
	%\rowcolor{lb}
	    $1/\epsilon$ & Learning rate  &         $\beta=g (1-2D\alpha\gamma)/2D\gamma\epsilon$ 	& Consumption rate of chemoattractant per cell\\
	\hline
	%\rowcolor{lb}
	Eq.~\eqref{eq:mf-risk} & Hamilton-Jacobi-Bellman equation	& Eq.~\eqref{eq:conc-ks} & Patlak-Keller-Segel equations \\
	\hline
\end{tabular}
\vspace*{0.2cm}
\caption{\textit{A dictionary between optimal cooperative search and chemotaxis.} \textmd{The table illustrates the correspondence between quantities in mean-field optimal control and their counterparts in chemotaxis.}}\label{tab:corr}
\end{table*}

\paragraph*{Optimal cooperative search.}
Let us consider a group of agents whose goal is to reach some target while minimizing a cost function that 
%We may think of the target as the exit from an enclosed domain, or the site of available resources. The cost 
is a sum over several contributions: time to reach the target, energy expenditure and a penalty for collisions.

The dynamics of the agents is given by a drift-diffusion equation
\begin{equation}\label{eq:cd}
%	d\XI_t = u^{(i)}(X^{(1)}_t \ldots X^{(N)}_t)\,dt + \sqrt{2 D}\,dW^{(i)}_t \ ,
	\frac{dX_i}{dt} = u_i + \sqrt{2D}\,\eta_i(t)
\end{equation}
where the subscript $i$ labels the agent, $X_i$ are the positions, $u_i$ are the individual controls, and $\eta_i$ are independent standard white noises, $\langle \eta_i(t)\,\eta_j(t')\rangle = \delta_{i,j}\delta(t-t')$. Uncontrolled motility is characterized by the constant diffusion coefficient $D$.
In general, the controls $u_i$ depend on the spatial configuration of all agents $X_1 \ldots X_N$.
The cost per unit time paid by the agent $i$ is
\begin{equation}\label{eq:cost}
	c_i = \underbrace{\vphantom{\sum_{j\neq i}}\ q(X_i)\ }_{\mbox{\footnotesize time}} + \underbrace{\vphantom{\sum_{j\neq i}}\ \frac{\gamma}{2}u_i^2\ }_{\mbox{\footnotesize energy}}  + \underbrace{\ \frac{g}{2}\sum_{j\neq i} \delta(X_i - X_j)\ }_{\mbox{\footnotesize  collisions}} \ .
\end{equation}
The total cost accumulated along the search process by the agent $i$ is the integral of the cost rate $c_i$ over time. When the agent reaches the target the cost does not increase anymore.
The cooperative search is completed when all the agents have reached the target.

The cost features three contributions. The first one is the penalty for the time spent before reaching the target. We denote the associated cost per unit time as $q$.
%, which may in general depend on the spatial location to account for spatial inhomogeneities. 
The second one is the cost of control, that we take as $\gamma u^2/2$. This is reminiscent of the power dissipation due to motion in a viscous medium, but can also be interpreted as the Kullback-Leibler divergence from a random strategy in the decision-making context \cite{todorov-09}. Other choices are possible that leave the scenario below largely unchanged (see SI, Sec. ). 
Finally, the last term arises from collisions. 
%In the simplest case, these occur upon contact with a strength specified by a parameter $g$. 
The general case of longer-range interactions between agents is discussed in the Supporting Information (SI, Section 1). This combination of factors embodies the trade-offs between different costs that lead to nontrivial solutions of the optimization problem: for instance, a fast search and a low collision risk cannot be achieved without a consequent expenditure in control cost.

The optimal control for the multi-agent system is the set of vector fields $u_i^*$ that minimize the total cost averaged over all the possible trajectories of the agents under the controlled dynamics
$\langle\sum_i \int c_i dt\rangle$ where the average is taken over the paths described by Eq.\,\eqref{eq:cd}.
This is the usual risk-neutral formulation of the optimal control problem, which corresponds to setting $\alpha=0$ in Table~\ref{tab:corr} --- see below for the risk-sensitive case.
See Fig.~1  in the SI for a schematic representation of the control problem.

The minimization of the cost functional can be performed by the Pontryagin minumum principle \cite{pontryagin-87}. This leads to the (non-linear) optimality Bellman equations, which can then be cast into a linear form by means of a Hopf--Cole transform \cite{todorov-09} (see Section 1.1 of {SI} for a full derivation).
The optimal control turns out to be
\begin{equation}\label{eq:oc-des}
	u_i^* = 2D\,\nabla_i\log {Z} \ ,
\end{equation}
where the desirability ${Z}(x_1\cdots x_N)$ %, known as desirability in the decision-making context, 
satisfies the linearized Bellman equation 
\begin{equation}\label{eq:hjb-des}
	%H\,{Z} \equiv 
	- D \sum_i \nabla^2_i {Z} + \frac{1}{2D\gamma}\sum_i h_i\, {Z}  = 0 \ ,
\end{equation}
where $h_i(x_1 \cdots x_N)$ is the contribution to the single-agent cost $c_i$ given by time expenditure and collisions, $h_i = q(x_i) + (g/2)\sum_{j\neq i}\delta(x_i-x_j)$.
The desirability ${Z}$ is a non-negative function of the configuration which is closely related to the optimal cost function ${C^*}(x_1\ldots x_N)$. The latter is defined as the minimum expected value of the total cost for an initial configuration $x_1 \ldots x_N$, which is achieved under the optimal control $u^*$. Explicitly, one has $Z = \exp[ - {C^*}/2D\gamma]$.
It is then clear that the optimal control biases the motion of the agents towards configurations with lower expected cost.
Eq.~\eqref{eq:hjb-des} has to be supplemented by appropriate boundary conditions, i.e. ${Z}=0$ for forbidden configurations, Neumann conditions on rigid walls, and more complicated ones at the target, which involve {the solutions of the control problem for any number of agents less than $N$ (see SI, Sec 1.1.1)}.
%Notice that the optimal controls for the individual particles generally depends on the positions of all the $N$ agents, because of the interaction term $v$. 
Note that Eq.~\eqref{eq:hjb-des} is equivalent to the stationary Schr\"odinger equation of a quantum-mechanical many-body system of identical particles \cite{lieb1-63, lieb2-63}.
To the best of our knowledge, an exact solution that satisfies the appropriate boundary conditions  is not known for a generic $N$, even for {simple geometries}. Moreover, a numerical approach appears to be a daunting task already for three agents in a two-dimensional domain. Approximation schemes are therefore very valuable in order to proceed further.

\paragraph*{Mean-field cooperative search and the emergence of chemotaxis.}
Guided by the interpretation of the linear Bellman equation as a quantum many-body problem of identical particles with short-range interaction, we adopted a mean-field approximation scheme, which is often successful in capturing the large-scale features of interacting systems \cite{parisi-88}. %,brezin-76}.
We remark in passing that the mean-field approach that we take here is exactly equivalent to the game-theoretical notion of cooperative mean-field game \cite{lasry-07} which has been applied to crowd dynamics in a fast evacuation scenario \cite{burger-13}. 

Mean-field solutions are in general suboptimal, since a certain amount of information is discarded by the agents in the evaluation of the optimal action. However, if $N$ is large and the system is diluted enough, a mean-field approximation for Eq.~\eqref{eq:hjb-des} yields an excellent approximation -- it actually becomes exact in the closely related problem of a confined, repulsive Bose gas \cite{lieb-01}.
We note in passing that when the agents are all identical the best mean-field solution is the cooperative one (see SI, Sec. 2.1.1).

This approximation consists in solving Eq.\,\eqref{eq:hjb-des} with the ansatz that the many-agent desirability $Z$ can be factorized in $N$ copies of a single function $\zeta$
\begin{equation}\label{eq:mf}
	Z(x_1\ldots x_N) = \zeta(x_1)\cdots\zeta(x_N) \ .
\end{equation}
In this ansatz, the control exerted by each agent is only determined by its own position $x$. Indeed, by combining Eqs.~\eqref{eq:oc-des} and~\eqref{eq:mf} it follows that
\begin{equation}\label{eq:mf-oc}
	u^* = 2D\nabla \log \zeta \ .
\end{equation}
The unique function $\zeta$, which can be read as the desirability of a spatial location for a single agent immersed in a crowd, satisfies the mean-field Bellman equation
\begin{equation}\label{eq:mf-des}
	D\nabla^2\zeta - \frac{1}{2D\gamma}\big( q + g(N-1)\rho \big)\,\zeta = 0  \ ,
\end{equation}
where $\rho$ is the single-agent probability density that obeys the Fokker-Planck equation
\begin{equation}\label{eq:fp}
	\partial_t \rho + \nabla \cdot ( \rho u^*) = D\nabla^2\rho \;.
\end{equation}
%The boundary conditions for the desirability are $\zeta=1$ at the target (no further cost for agents which are already on target), $\zeta=0$ in forbidden regions (infinite cost) and zero-flux at rigid walls.

%Comparing the exact Eq.~\eqref{eq:hjb-des} with its mean-field approximation Eq.~\eqref{eq:mf-des} one sees that the multi-agent ``potential'' has been replaced by an effective single-agent potential which depends on the probability density, as usual in mean-field approaches (see Methods).

Remarkably, the set of equations \eqref{eq:mf-oc}, \eqref{eq:mf-des} and \eqref{eq:fp} is identical, within proportionality factors, to a limiting case of the well-known Patlak--Keller--Segel equations, which was first introduced to model microbial chemotaxis at the population level \cite{patlak-53,ks-71}
\begin{equation}\label{eq:conc-ks}
\begin{array}{l}
\partial_t n + \nabla \cdot ( \chi n \nabla \log s) = D\nabla^2 n \,, \\
	D_s\nabla^2 s - k s - \beta n  s  = 0 \;.
\end{array}
\end{equation}
where $n$ is the number density of microbes, $s$ is the concentration of chemoattractant and $D_s$ is its molecular diffusivity.
Comparing the Bellman equation Eq.~\eqref{eq:mf-des} with the second row of Eq.~\eqref{eq:conc-ks} one sees that the desirability $\zeta$ is proportional to the chemoattractant concentration $c$,
to which agents respond logarithmically --  they sense only fold-changes in levels, in accord with the Weber--Fechner law \cite{adler-17}. The chemotactic coefficient is $\chi=2D$ in this case.
The chemoattractant is degraded with rate $k$ proportional to $q/(2 D \gamma)$ and consumed at rate $\beta$ proportional to $g/(2D \gamma)$ per cell.
We note in passing that perfect adaptation is an implicit feature of Eqs.\,\eqref{eq:mf-oc} and \eqref{eq:fp}, in that there is no chemokinesis -- random motility $D$ does not depend on $\zeta$.

\paragraph*{Learning to search optimally: scouts, beacons and recruitment.}
The main results of the previous sections are that optimal cooperative search can be realized by biophysical systems in which the target emits a diffusible chemical cue in the environment, that agents respond chemotactically to this signal, and actively modify it. However, it would be useful to extend this setting to the relevant case when the location of the target is a priori unknown and the target does not spontaneously send out signals to facilitate the work of the agents. In other words, we seek a way to include the process of discovery of the target's location and the successive construction of the solution to Eq.~\eqref{eq:mf-des}. As we show below, this can be accomplished by adding a production term in the equation for concentration, which is the exact analog of the process by which information is written on memory.  
%The resulting strategy provides an optimal solution to problems like collective predation or crowd evacuation, for example.

Our solution to the learning problem goes as follows. Initially, the concentration is set to a constant everywhere in space. In the first part of the search process, agents wander at random and the concentration decays and is consumed. As a result, agents explore space away from their initial location. This is called the scouting process. 
When agents eventually reach the target, they start the production of chemoattractant on site, either releasing it themselves, e.g. in the form of a pheromone-like cue \cite{wyatt-03, beacon-10},
or inducing its production by the target, which may happen in practice by triggering specific gene expression \cite{humphreys-12} or by transforming it into attractive waste material. 
The net effect is that a beacon signal is emitted from the target, and it leads to the recruitment of all other agents towards it. 
%Similar concepts have been exploited in biomimetic applications such as collaborative foraging by robots \cite{beacon-10}.
A mathematically precise description of  the process outlined above requires only the addition of two terms to the optimality equation \eqref{eq:mf-des}
\begin{equation}\label{eq:mf-learning}
	\underbrace{\epsilon\partial_t \zeta}_{\mbox{\footnotesize relaxation}} -D\nabla^2\zeta + \frac{1}{2D\gamma}\big( q + g(N-1)\rho \big)\,\zeta =  \underbrace{f(t) \mathbbm{1}_{\tt target}}_{\mbox{\footnotesize production}}  \ ,
\end{equation}
where  $f(t)=\int_0^t dt'\,\int_{\tt target} ds\cdot J_\rho $ is the cumulated number of agents which have reached the target up to time $t$,  and $J_\rho = (2D\nabla \log \zeta) \rho - D \nabla \rho$ is the spatial flux of agents. The indicator function $\mathbbm{1}_{\tt target}$ specifies that production takes place only at the target. The relaxation term is interpreted as a delay in writing information on memory, i.e. a learning rate. When production and diffusion balance, the optimal solution, Eq.\,\eqref{eq:mf-des}, is attained. 
Notice that $\epsilon$ is the proportionality factor between  Eq.\,\eqref{eq:mf-des} and the second equation in \eqref{eq:conc-ks}.

\begin{figure}[b]
	\centering
	\includegraphics[width=\columnwidth]{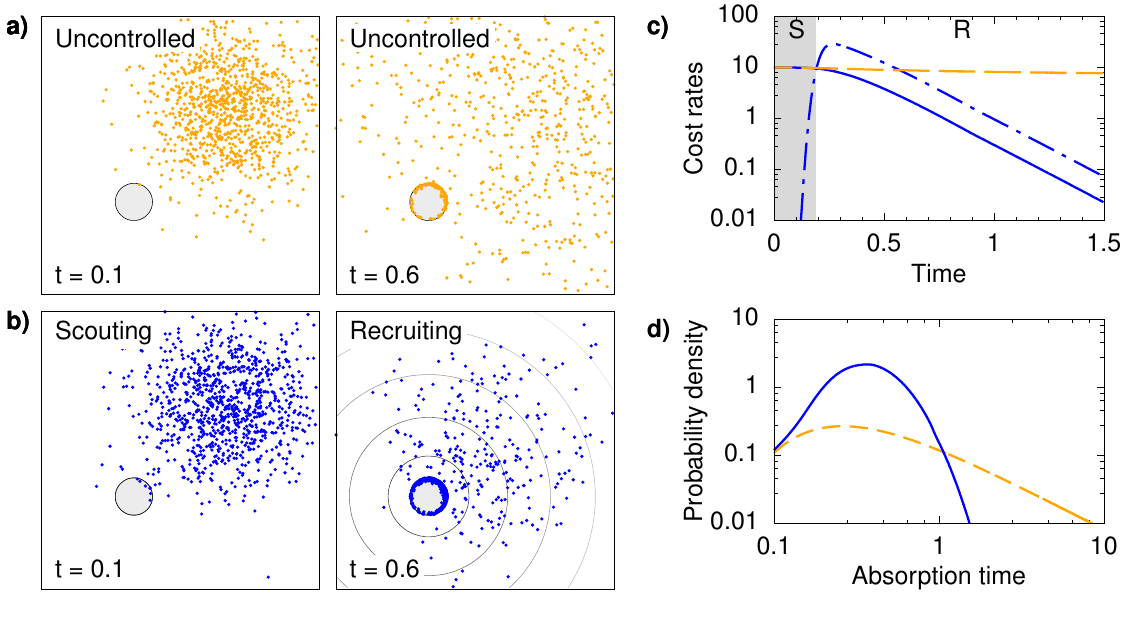}
	\caption{\textit{Optimal cooperative predation.}
		Comparison between the uncontrolled \textit{a)} and controlled dynamics \textit{b)}, in the non-interacting case ($g=0$). The agents are initially localized in a small region of space and are required to reach the target (grey disk).
				They initially undergo unbiased diffusion during the scouting phase
				%. In this stage, the controlled dynamics is essentially equivalent to the uncontrolled one because of the absence of collision costs.
		and when some reach the target, the recruiting phase begins. The chemical cue is emitted from the target and degraded at constant rate, resulting in a gradient (grey contour lines, in logarithmic scale) which elicits a drift toward the target in all other agents. In these simulations the parameters are: $\gamma = 1$, $q = 10$, $D=1$, $g=0$, $\epsilon=0.1$.
		\textit{c)}: Average cost rate for time (uncontrolled: dashed orange line; controlled: solid blue line) and for control (dash-dotted blue line)
		% are shown -- proportional to the fraction of agents which have not reached the target yet: the solid blue line represents the cost for time for the controlled agents, the dashed orange stands for the same quantity for the uncontrolled ones, while the dashed-dotted blue line represent the average cost rate for the control.		
		The scouting phase (S, shaded) and the recruiting (R) phase are dominated by time cost and by control cost, respectively.
		\textit{d)}: Probability density function of the time to reach the target (color code as in \textit{c}). For small times the distributions are similar, while at large times controlled agents display an exponential decay against a $-3/2$ power law	for uncontrolled ones.}\label{fig:circle}
\end{figure}

In the remainder of this section we illustrate how the optimal solution is achieved in two examples of cooperative search games. The first example features a circular target in a two-dimensional domain and 
can be thought of as a basic model for bacterial predation \cite{perez-16, humphreys-12}.
In Fig.\,\ref{fig:circle} we show the simulation of a large number of agents under the controlled dynamics with optimal mean-field drift, which can be computed exactly in this case (see SI, Sec. 3), and compare them with the uncontrolled dynamics.
From visual inspection, the gain in the number of agents that have reached the target is apparent.
{More quantitatively, the time average cost per agent as a function of time (Panels c and d of Fig.\,\ref{fig:circle}),
which is proportional to the number of agents which have not reached the target at a given time, falls off exponentially for the controlled case while it exhibits a very slow decay for uncontrolled diffusion.}
%
%\begin{figure*}[t!]
%\end{figure*}

\begin{figure}[t]
	\centering
	\includegraphics[width=\columnwidth]{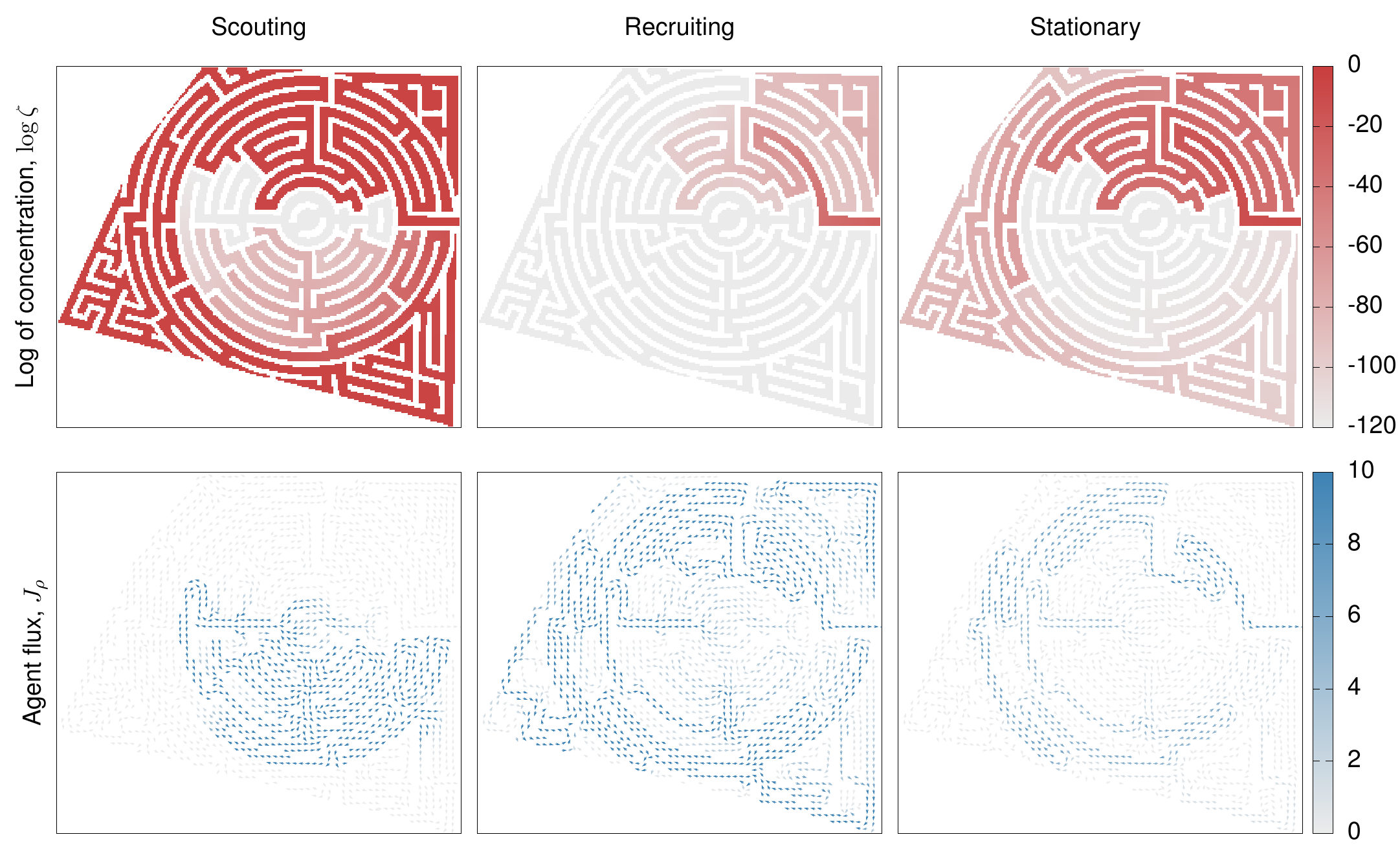}
	\caption{\textit{Optimal crowd evacuation.}
		Agents are injected at a constant rate at the center of the maze and have to find the exit (on the right side of the maze), as quickly as possible while minimizing collisions.
				The panels show a numerical simulation of Eqs.~\eqref{eq:mf-oc}, \eqref{eq:fp} and \eqref{eq:mf-learning}. The desirability (=concentration, see Table~\ref{tab:corr}) is shown in the top panels, while the flux of agents $J_\rho$ is displayed in the bottom panels. During scouting (left column), the population consumes the chemical, leading to an outward-driven scouting process, faster than pure diffusion. Upon reaching the target, agents lay the beacon signal and recruit those who lag behind to the target (middle column). Eventually, since agents are continuously injected in this case, a stationary state is reached where agents track the optimal path from the center to the exit (right column).
The parameters are $D=1$, $\gamma = 1$, $\epsilon = 10^{-1}$, $q=10$ and $g(N-1) = 100$.	}
	\label{fig:maze-n}
\end{figure}

The second example is crowd evacuation from a complicated domain.
Agents, initially localized in the center of a maze, are required to find the exit with the minimal cost.
The domain in which we performed this numerical experiment is a reproduction of the historical maze in the gardens of Villa Pisani (Stra, Italy). % depicted in Fig. \ref{fig:pisani}.
In this example, agents are introduced at the center of the maze at a constant injection rate. % (corresponding to the tower of the picture of Fig. \ref{fig:pisani}.
In Fig. \ref{fig:maze-n} we see the emergence of the phases of scouting and recruitment, and eventually, we observe that the agents trace out the optimal path to the exit.
Notice that, during the scouting phase, the density of agents propagates as a front with speed which is proportional to $\sqrt{N}$ (see SI, Sec. 2.1.2) so that the collective is much faster in finding the target than the single agent (which instead reaches it diffusively).

\paragraph*{Extension to risk-sensitive control.}
%In the previous sections we have considered the problem of minimizing the average cost, the so-called risk-neutral case.
%In a more general framework, it might be of interest to minimize different statistics of the cost. For example, one may wish to limit the effect of improbable but very expensive events by means of a risk-averse strategy.
%Does the optimal mean-field solution correspond to chemotaxis in this case as well? How is the dictionary affected by risk-sensitivity?

A convenient way of incorporating the notion of risk in decision making is to introduce a parameter $\alpha$ which exponentially weighs the fluctuations of the cost \cite{howard-72,dvijotham-11}. In this setting the functional to be minimized becomes $\mathcal{F}_\alpha=\alpha^{-1}\log\langle \exp (\alpha\sum_i \int c_i dt) \rangle$ ({see SI for details}). This choice ensures the invariance of the optimal control under a global offset of the costs.
It is easy to verify that 
as $\alpha \to 0$ one recovers  the risk-neutral case previously described.
When $\alpha$ is positive the optimal control is such that fluctuations with cost higher than average are suppressed, and one refers to it as a risk-averse controller. Conversely, 
when $\alpha$ is negative, the optimal controller feels optimistic and is risk-seeking. In this case low-cost fluctuations are enhanced. The procedure described in the previous section, including the mean-field approximation, can be extended to the risk-sensitive setting (see Sections 1.2 and 2.2 of SI for a full derivation). 

We find that the optimal solution to the risk-sensitive cooperative search game is also described by the chemotaxis equations, and a direct comparison between Eq.\,\eqref{eq:mf-risk} (with the addition of learning) and Eq.\,\eqref{eq:conc-ks} yields the dictionary in Table~\ref{tab:corr}. The risk-sensitive optimality equations
%formally obtained from the risk-neutral ones by a simple rescaling of the chemotactic coefficient and the degradation rates.
that generalize  Eqs.\,\eqref{eq:mf-oc} and \eqref{eq:mf-des} are
\begin{equation}\label{eq:mf-risk}
	\left\{
	\begin{aligned}
		&u^* = \frac{2D}{1-2D\alpha\gamma} \nabla \log\zeta \ , \\
		&D\nabla^2\zeta - \frac{1-2D\alpha\gamma}{2D\gamma}\big(q + g(N-1) \rho\big)\,\zeta = 0\, ,
	\end{aligned}
	\right.
\end{equation}
where $2D\alpha\gamma<1$.

\paragraph*{Discussion.}
%We have shown how the problem of finding the optimal solution to a cooperative search game can be solved by agents which produce and consume a diffusible chemical cue, establishing a close correspondence between notions from decision-making theory and biophysical properties of chemotactic microorganisms. What are the implications of our results? 

From the standpoint of search theory, our findings provide a solid theoretical rationale for the many solution methods inspired by chemotaxis, from computational \cite{passino-02,muller-02,reynolds-10}, to biological \cite{nakagaki-00, nakagaki-07} and physico-chemical ones \cite{lagzi-10,jin-17}. At the practical level, we offer explicit expressions for the optimal choice for the parameters that appear in these biomimetic approaches, allowing to shortcut the painstaking procedure of parameter tuning. 
%When the exact optimal values are difficult to realize in practice, as it may happen for real biological or physico-chemical systems, our analysis permits to evaluate the impact of the suboptimal choice on the performance of the search method.
Conversely, from the viewpoint of chemotaxis, we remark that the dictionary in Table~\ref{tab:corr} can also be read in reverse, which allows to solve the inverse problem of retrieving the decision-making parameters from biophysical observations. 
%Indeed, from the experimental knowledge of the effective microbial diffusivity $D$, the chemotactic coefficient $\chi$, the chemoattractant diffusivity $D_s$, the degradation rate $k$ and consumption rate $\beta$, one can invert the equations in Table~\ref{tab:corr} and extract the values of  $\alpha$, $\gamma$, $q$ and $g$ to within an overall, irrelevant multiplicative factor. 
For example, bacterial chemotaxis experiments (Fig.~6 in Ref.~\cite{kalinin-09}) give $\chi/D \approx 12$, which translates into $2 D\alpha\gamma \approx 5/6$. This value is very close to the upper limit for risk aversion, suggesting that that bacteria try to minimize the impact of unfavorable fluctuations -- a conclusion that has also been reached by other means~\cite{celani-10}.  

%Our results open many questions and directions for further research. Among them, extending the present analysis to cooperative Markov games \cite{littman-94} would allow to address problems of traffic control on arbitrary networks, for example, as well as many problems of cooperative resource allocation and transfer \cite{hohzaki-16}. What will be the equivalent of chemical communication in these contexts? Another exciting direction to investigate is the realm of non-cooperative games and ecology: will optimal strategies and population dynamics be shaped by chemical signalling in this case as well?

\paragraph*{Acknowledgments.}
AC acknowledges innumerable inspiring discussions with Massimo Vergassola. We are grateful to Rami Pugatch for useful comments and suggestions.

%\bibliography{References}
\bibliography{shorttitles,shortRefs}

%\pagebreak
\newpage
\pagestyle{empty}


\section*{Supplementary material}

# Supplemental Information

## 1. DERIVATION OF THE BELLMAN EQUATION FOR OPTIMAL COOPERATIVE SEARCH

Let us consider a system of $N$ agents in $d$ dimensions[1] following the stochastic dynamics [2]

$$d\bar{X}^t = \bar{u}(\bar{X}^t, t) \, dt + \sqrt{2D} \, d\bar{W}^t \, , \tag{1}$$

where $\bar{W}$ is the standard vector Wiener process, representing the uncontrolled dynamics, and $\bar{u}$ is the *control*. In Eq. (1) $D$ is the diffusion coefficient, which we choose to be constant. Notice that in this case there is no ambiguity about the regularization, in that Ito or Stratonovich conventions are equivalent. Part of the domain boundary is absorbing and we refer to it as the *target*. We define the cost functional as

$$C_0^T = \sum_{i=1}^{N} \int_0^{\min\{T, T_i\}} dt \left( \frac{\gamma}{2} u_i(\bar{X}^t, t)^2 + h_i(\bar{X}^t) \right) \, , \tag{2}$$

where

$$h_i(\bar{X}^t) = q(X_i^t) + \frac{g}{2} \sum_{j \neq i} v(X_i^t, X_j^t) \, ,$$

$q > 0$ and $g > 0$, $\gamma > 0$ and $T_i$ is the exit time (first passage at the target) for the $i$-th agent. The upper extreme of integration in time indicates that an agent stops contributing to the total cost as soon as it hits the target. The quadratic form of the cost for the control is directly related to the Kullback–Leibler divergence of the controlled path measure from the uncontrolled (pure diffusion) one; therefore, the cost of the control has a natural probabilistic interpretation in terms of "distance" between path measures [1].

We first discuss the finite-horizon setup, in which the objective of the optimization is a functional of the cost accumulated over the fixed interval of time $[0, T]$, $C_0^T$. In this case the problem is generally time-dependent. The finite horizon setup is more general than the terminal state setup discussed in the main text, which essentially corresponds to the limit $T \to \infty$ of Eq. (2), i.e. infinite horizon limit in presence of terminal states. In Fig. 1 is shown a schematic representation of the terminal state problem defined in the main text, where $v(x_1, x_2) = g\delta(x_1 - x_2)$.

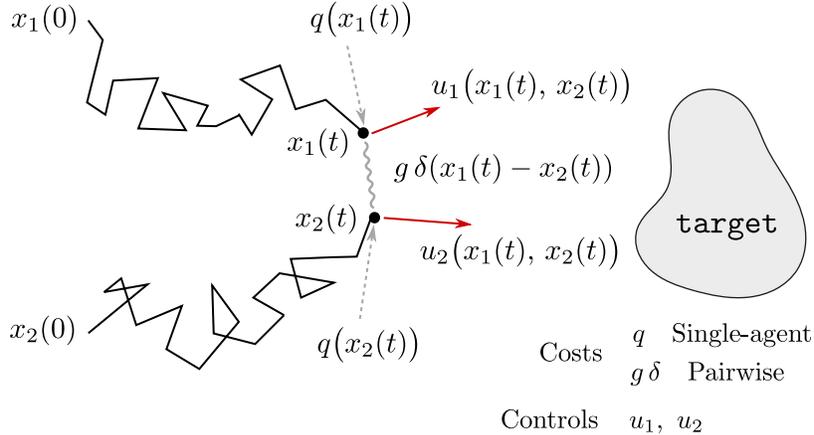

FIG. 1. *Scheme of the cooperative search game for $N = 2$ agents.* The agents are driven by the controls $u_1$ and $u_2$, which depend on both their positions $x_1$ and $x_2$. They pay individual costs associated with control and time expenditure, respectively with a rate $\gamma u^2/2$ and $q$ (possibly position-dependent), and a pairwise cost associated with the interaction – collision – with a rate proportional to a Dirac-$\delta$ function. Optimal controls minimize the average cost (or an exponential average in the risk-sensitive case).

---

[1] In the context of decision making, this is the term we use referring to controlled particles. Indeed, controlled diffusion is a limit of a Markov decision process in which the term "agent" is more natural. In this Supporting Information notes we will use the words "agent" and "particle" interchangeably.

[2] Symbols expressed with a *bar* indicate $N$-tuples whose index corresponds to the label of the agent; e.g. $\bar{x} \equiv \{x_1 \dots x_N\}$. The superscript $t$ indicates time.

### 1.1. Risk-neutral case

Here we derive the Bellman equation for the optimal control which minimizes the *average* of the $N$-particle cost function (2). In terms of the $N$-particle density function, $p$, the average cost is expressed as

$$\mathcal{F}[\bar{u}] = \langle C_0^T \rangle$$
$$= \int_0^T dt \int d^N x \sum_i \left( \frac{\gamma}{2} u_i(\bar{x}, t)^2 + h_i(\bar{x}) \right) p(\bar{x}, t) \ . \tag{3}$$

Since $p$ has an implicit dependence on $u$ through the Fokker–Planck equation associated with Eq. (1), it is convenient to couch the minimization problem by including the dynamics as a constraint and seek to minimize the auxiliary functional

$$\mathcal{L}[p, \bar{u}, \Phi] = \mathcal{F} + \int_0^T dt \int d^N x \, \Phi(\bar{x}, t) \left[ \partial_t p(\bar{x}, t) + \sum_i \nabla_i \cdot \left( u_i(\bar{x}, t) p(\bar{x}, t) - D \nabla_i p(\bar{x}, t) \right) \right] \ , \tag{4}$$

where $\Phi(\bar{x}, t)$ is a Lagrange multiplier. This is an application of the so-called Pontryagin minimum principle [2]. The condition of null variation of $\mathcal{L}$ with respect to $\Phi$ yields the Fokker–Planck equation for the $N$-particle density $p$. The variation of $\mathcal{L}$ with respect to $u_i$, at the saddle point,

$$\left. \frac{\delta \mathcal{L}}{\delta u_i} \right|_* = \left( \gamma \, u_i^* - \nabla_i \Phi \right) p = 0 \ , \tag{5}$$

gives the optimal control

$$u_i^*(\bar{x}, t) = \gamma^{-1} \nabla_i \Phi(\bar{x}, t) \ . \tag{6}$$

Finally, the variation with respect to $p$ gives

$$\left. \frac{\delta \mathcal{L}}{\delta p} \right|_* = -\partial_t \Phi + \sum_i \left( \frac{\gamma}{2} (u_i^*)^2 - u_i^* \cdot \nabla_i \Phi - D \nabla_i^2 \Phi + h_i \right) = 0 \ , \tag{7}$$

which, together with Eq. (6), leads to the Hamilton–Jacobi–Bellman (HJB) equation:

$$\partial_t \Phi = -\frac{1}{2\gamma} \sum_i \left( \nabla_i \Phi \right)^2 - D \sum_i \nabla_i^2 \Phi + \sum_i h_i \ . \tag{8}$$

As pointed out in the main text, the function $\Phi(\bar{x}, t)$ is the optimal value function at time $t$ and in the state $\bar{x}$, up to an additive constant: this is (minus) the expected cost-to-go under the optimal control $\bar{u}^*$ when the system is conditioned to be in state $\bar{x}$ at time $t$, namely

$$\Phi(\bar{x}, t) \equiv -\left\langle C_t^T \,\middle|\, \bar{X}^t = \bar{x} \right\rangle_* = -\int_t^\infty dt' \int d^N x' \sum_i \left( \frac{\gamma}{2} u_i^*(\bar{x}', t')^2 + h_i(\bar{x}') \right) p(\bar{x}', t') \ , \tag{9}$$

where $p$ satisfies the Fokker–Planck equation with control $\bar{u}^*$ and has initial condition

$$p(\bar{x}', t) = \delta^N(\bar{x}' - \bar{x}) \ .$$

Indeed, it can be directly verified that the r.h.s. of Eq. (9) satisfies the saddle-point equation (7).

The HJB equation (8) is non-linear. However, it is possible to make it linear by the Hopf–Cole transformation,

$$Z = \exp\left( \Phi / 2D\gamma \right) \tag{10}$$

which turns Eq. (8) into

$$\partial_t Z = -D \sum_i \nabla_i^2 Z + \frac{1}{2D\gamma} \, q \, Z \ , \tag{11}$$

and the optimal control into

$$u_i^* = 2D \nabla_i \log Z \ . \tag{12}$$

In the main text we focused on a terminal state optimal control problem. In such a setup, when the uncontrolled dynamics is time-homogeneous and the costs $h_i$ do not depend explicitly on time, the HJB equation admits stationary solutions, $\Phi(\bar{x})$ and $Z(\bar{x})$.

### 1.1.1. Boundary conditions

The optimal control is found from the stationary solution of Eq. (11) with appropriate boundary conditions. Here we discuss the structure of these boundary conditions at the target. We start by recalling the relationship of the desirability $Z$ with the optimal cost-to-go function,

$$Z(\bar{x}) = e^{-\langle C_0^\infty | \bar{X}^0 = \bar{x} \rangle_* / 2D\gamma} \ ,$$

obtained from Eq. (9) and the Hopf–Cole transformation. From the definition of the cost function, once an agent reaches the target it does not contribute to the cost any longer. Let us assume that all the agents are inside the domain except one, which we choose to be agent $N$ without loss of generality, which sits at the target: $X_N^0 = x_N \in \mathtt{target}$. Then, since agent $N$ does not contribute to the cost, the desirability is a function of the remaining $N-1$ agents only [3]

$$Z^{(N)}(x_1 \ldots x_{N-1}, \, x_N)\big|_{x_N \in \mathtt{target}} = Z^{(N-1)}(x_1 \ldots x_{N-1}) \ ; \tag{13}$$

more generally, if the agents $1 \ldots i$ (up to relabeling) are not yet at the target, while the others have already reached it, one has

$$Z^{(N)}(x_1 \ldots x_i, \, x_{i+1} \ldots x_N)\big|_{x_{i+1} \ldots x_N \in \mathtt{target}} = Z^{(i)}(x_1 \ldots x_i) \ . \tag{14}$$

## 1.2. Risk-sensitive case

Here we show that the risk-sensitive Hamilton-Jacobi-Bellman equation can also be linearized. In the risk-sensitive setup, agents aim at the minimization of the average of an exponentially weighted cost [3]

$$\mathcal{F}_\alpha = \frac{1}{\alpha} \log \langle e^{\alpha C_0^T} \rangle = \frac{1}{\alpha} \log \mathcal{G}_\alpha \ , \tag{15}$$

where $C_0^T$ is defined in Eq. (2). In this section we only consider $\alpha > 0$, so the problem is equivalent to the minimization of the functional $\mathcal{G}_\alpha = \exp \alpha \mathcal{F}_\alpha$. It is easy to generalize the derivation for $\alpha < 0$.

The parameter $\alpha$ defines the *risk sensitivity* of the optimal control problem: one recognizes that in the limit $\alpha \to 0$, $\mathcal{F}_\alpha \to \mathcal{F} = \langle C_0^T \rangle$, and the control problem is know as *risk-neutral*; if $\alpha > 0$ the optimal solution is the one that reduces the most the fluctuations towards high values of the cost, corresponding to a *risk-averse* strategy; finally, $\alpha < 0$ corresponds to an optimization in which more weight is given to the values of $C_0^T$ which are smaller than average, therefore leading to *risk-seeking* strategies. The limits $\alpha \to \infty$ and $\alpha \to -\infty$ are referred to as *min-max* and *min-min* optimization problems, respectively.

We apply the Pontryagin principle to the minimization of the functional $\mathcal{G}_\alpha$. It is convenient to consider the Markov process $\{\bar{X}^t, \, C_0^t\}_t$, where $\bar{X}^t$ follows the stochastic dynamics in Eq. (1), while from Eq. (2)

$$dC_0^t = \sum_i \left( \frac{\gamma}{2} u_i(\bar{X}^t, t)^2 + h_i(\bar{X}^t) \right) dt \equiv c(\bar{X}^t) \, dt$$

The Fokker–Planck equation associated to this system of equations, which describes the evolution of the probability density function $p(\bar{x}, C, t)$ for the process $\{\bar{X}^t, C_0^t\}$, is

$$\partial_t p + \partial_C \big( c(\bar{x}) \, p \big) + \sum_i \nabla_i \big( u_i \, p - D \nabla_i p \big) = 0 \tag{16}$$

The functional $\mathcal{G}_\alpha$ can be then expressed as a linear functional of $p$:

$$\mathcal{G}_\alpha = \int d^N x \, dC \, p(\bar{x}, C, T) \, e^{\alpha C} \equiv \int d^N x \, G_\alpha(\bar{x}, T) \ ; \tag{17}$$

the function $G_\alpha(\bar{x}, t)$ is the average of $\exp \alpha C_0^t$ over all trajectories which have arrived at $\bar{x}$ at time $t$. By multiplying Eq. (16) by $\exp \alpha C$ and integrating over $C$, one recovers the forward Feynman–Kac equation for $G_\alpha$:

$$\partial_t G_\alpha + \sum_i \nabla_i \cdot \left( u_i G_\alpha - D \nabla_i G_\alpha \right) = \alpha \, c \, G_\alpha \ . \tag{18}$$

---

[3] For the sole purpose of illustrating the boundary conditions defining the many-particle optimal control problem, we introduce the notation $Z^{(n)}$ to indicate the desirability function $Z$ for the problem with $n$ agents (when $Z$ is a function of $n$ spatial variables).

The Pontryagin principle is then applied to the minimization of the functional $\mathcal{F}_\alpha$ subject to the constraint imposed by Eq. (18); this is equivalent to perform the unconstrained minimization of the Lagrange functional

$$\mathcal{L}_\alpha = \underbrace{\int d^N x\, G_\alpha(\bar{x}, T)}_{\mathcal{G}_\alpha} + \int_0^T dt \int d^N x\, \Psi_\alpha(\bar{x}, t) \Big[ \partial_t G_\alpha + \sum_i \nabla_i \cdot \big( u_i G_\alpha - D \nabla_i G_\alpha \big) - \alpha \sum_i \Big( \frac{\gamma}{2} u_i^2 + h_i \Big) G_\alpha \Big] \ . \quad (19)$$

At the saddle point, variation with respect to the control $u_i$ yields

$$\frac{\delta \mathcal{L}_\alpha}{\delta u_i}\Big|_* = -G_\alpha \Big( \nabla_i \Psi_\alpha + \alpha \gamma\, u_i^* \Psi_\alpha \Big) = 0 \quad (20)$$

so the optimal control is

$$u_i^* = -\frac{1}{\gamma \alpha} \nabla_i \log \Psi_\alpha \ . \quad (21)$$

The variation with respect to $G_\alpha$ gives

$$\frac{\partial \mathcal{L}_\alpha}{\delta G_\alpha}\Big|_* = \delta(t - T) - \partial_t \Psi_\alpha - \sum_i u_i^* \cdot \nabla_i \Psi_\alpha - D \sum_i \nabla_i^2 \Psi_\alpha - \alpha \sum_i \Big( \frac{\gamma}{2} u_i^{*2} + h_i \Big) \Psi_\alpha = 0 \ , \quad (22)$$

which is the backward Feynman–Kac equation for the function $\Psi_\alpha$, which can be interpreted (up to multiplicative constants) as

$$\Psi_\alpha(\bar{x}, t) \equiv \big\langle e^{\alpha C_t^T} \big| \bar{X}^t = \bar{x} \big\rangle \ ; \quad (23)$$

the $\delta$-function in time sets the condition at the final time, if a finite-horizon problem is considered; in our terminal state setup, $T \to \infty$ and the $\delta$-function disappears, so we omit it in the following. Note that $\Phi_\alpha \equiv -\alpha^{-1} \log \Psi_\alpha$, plays the role of the value, in that $u_i^* = \gamma^{-1} \nabla_i \Phi_\alpha$. Indeed, in the limit $\alpha \to 0$, $\Phi_\alpha$ reduces to the (risk-neutral) value function $\Phi$ [4]. We therefore identify $\Phi_\alpha$ with the *risk-sensitive value function* [4, 5]. The HJB equation for the risk-sensitive value function then is

$$\partial_t \Phi_\alpha = -\underbrace{\Big( \frac{1}{2\gamma} - D\alpha \Big)}_{\equiv 1/2\tilde{\gamma}} (\nabla_i \Phi_\alpha)^2 - D \sum_i \nabla_i^2 \Phi_\alpha + \sum_i h_i \ ; \quad (24)$$

this equation has the same form as Eq. (8), where the parameter $\gamma$ is replaced by $\tilde{\gamma} \equiv \gamma/(1 - 2D\alpha\gamma)$. In the same way as for the risk-neutral case, a linearized version of the HJB equation can be found by performing the Hopf–Cole transformation

$$Z_\alpha = \exp\left( \Phi_\alpha / 2D\tilde{\gamma} \right) \quad (25)$$

to obtain

$$\partial_t Z_\alpha = -D \sum_i \nabla_i^2 Z_\alpha + \frac{1}{2D\tilde{\gamma}} \sum_i h_i\, Z_\alpha \ . \quad (26)$$

The optimal control is hence obtained from $Z_\alpha$ as

$$u_i^* = \frac{2D\tilde{\gamma}}{\gamma} \nabla_i \log Z_\alpha \ . \quad (27)$$

It is straightforward that for $\alpha = 0$, the optimal control equations (26) and (27) exactly reduce to the risk-neutral ones, respectively (11) and (12). As for the boundary conditions for Eq. (26), the same considerations made for the risk-neutral case hold here: Eqs. (13) and (14) are valid also in this case, with $Z$ being replaced by $Z_\alpha$.

---

[4] $\Psi_\alpha$, as defined in Eq. (23), is the moment generating function for the statistics of the cost $C_t^T$ conditioned to $\bar{X}^t = \bar{x}$. When $\alpha \to 0$,

$$\Phi_\alpha = -\alpha^{-1} \log \Psi_\alpha \to -\frac{\partial \Psi_\alpha}{\partial \alpha}\Big|_{\alpha=0} = -\langle C_t^T | \bar{X}^t = \bar{x} \rangle \equiv \Phi \ .$$

### 1.3.   Exact solution for the 1D non-interacting case

In this subsection we offer the exact analytic calculation for the non-interacting search in one dimension. The results also provide an approximation to the solution for the multi-dimensional case at large distances from the target.

A single agent is initially at $x$ on the real axis and the target is at the origin; the generating function of the cost under the control $u$, $\tilde{G}_s(x) = \langle \exp(-sC_0^T)|X^0 = x\rangle$, satisfies the (stationary) Feynman–Kac equation

$$u\,\tilde{G}'_s + D\,\tilde{G}''_s = s\left(\frac{\gamma}{2}u^2 + q\right)\tilde{G}_s\;,\tag{28}$$

with boundary conditions $\tilde{G}_s(0) = 1$ and $\tilde{G}_s(\pm\infty) = 0$ if $s > 0$ and $\tilde{G}_s(\pm\infty) = +\infty$ if $s < 0$. Assuming that the control is constant and pointing toward the origin [5], $u = -\text{sign}(x)\,U$, with $U$ positive, one can easily find that Eq. (28) is solved by

$$\tilde{G}_s = \exp\left\{\beta|x|\left(1 - \sqrt{1+\Gamma s}\right)\right\}\;,\tag{29}$$

where $\beta = U/(2D)$ and $\Gamma = 2D(\gamma + 2q/U^2)$. We can indeed check that the optimal control $U^*$ obtained from the one-dimensional HJB equation is the one which minimizes $\mathcal{F}_\alpha|_{X^0=x} = \alpha^{-1}\log\tilde{G}_{-\alpha}$:

$$\frac{\partial}{\partial U}\mathcal{F}_\alpha|_{X^0=x} = \frac{\partial}{\partial U}\left\{\frac{U|x|}{2D\alpha}\left(1 - \sqrt{1 - 2D\alpha\left(\gamma + \frac{2q}{U^2}\right)}\right)\right\} = 0\;,\tag{30}$$

solved by

$$U^* = \sqrt{\frac{2q}{\gamma(1-2D\alpha\gamma)}}\;.\tag{31}$$

The probability distribution of the cost $C_0^T$ under the control $U$ is found by applying the inverse Laplace transform to Eq. (29),

$$p(c|X^0 = x) = \frac{1}{2\pi\mathrm{i}}\int_{-\mathrm{i}\infty}^{+\mathrm{i}\infty} ds\,\tilde{G}_s = \frac{e^{\beta|x|-c/\Gamma}}{\beta^2 x^2\Gamma}\,\frac{1}{2\pi\mathrm{i}}\int_{0^+-\mathrm{i}\infty}^{0^++\mathrm{i}\infty} dt\,e^{-\sqrt{t}}\,e^{t\,c/(\beta^2 x^2\Gamma)}$$

$$= \frac{\beta|x|\sqrt{\Gamma}\,e^{\beta|x|}}{2\sqrt{\pi}}\,c^{-3/2}\,e^{-(\beta^2 x^2\Gamma)/(4c)-c/\Gamma}\;,\tag{32}$$

To obtain the second equality the change of variable $t = (\beta x)^2(1+\Gamma s)$ has been made, while in the last equality one makes use of Eq. [31[8]] from Ref. [6]. The important remark is that in Eq. (32) the right tail of the probability density of the cost has an exponential cutoff with rate

$$\frac{1}{\Gamma} = \frac{1}{2D(\gamma + 2q/U^2)} < \frac{1}{2D\gamma} = \alpha_{\max}\;.\tag{33}$$

This result implies that, for any value of $U$, $\langle e^{\alpha C_0^T}\rangle$ diverges when $\alpha > \alpha_{\max}$. In particular, in the limit $\alpha \to \alpha_{\max}^-$, the functional $\langle e^{\alpha C_0^T}\rangle$ diverges also for controls arbitrarily close to the optimal one, for which $\Gamma^{-1} = \alpha_{\max}$.

### 1.4.   Robustness of the optimal solution

The analytical solution in one dimension also allows to address the question about the robustness of the cost against perturbations in the control away from optimality. For controls $U$ close to the optimal value $U^*$, the risk-sensitive cost $\mathcal{F}_\alpha$ can be approximated by a quadratic function,

$$\underbrace{\mathcal{F}_\alpha - \mathcal{F}_\alpha^*}_{\delta\mathcal{F}_\alpha} \simeq \frac{1}{2}\mathcal{F}''_\alpha(U^*)\,\big(\underbrace{U - U^*}_{\delta U}\big)^2\;.$$

---

[5] We know from the exact solution that the optimal control in one dimension is constant: this follows from the fact that the solution of the HJB equation $D\nabla^2 Z_\alpha = q/(2D\tilde{\gamma})\,Z_\alpha$ with the boundary conditions specified above is solved by $Z_\alpha = \exp\{-[q/(2D^2\tilde{\gamma})]^{1/2}|x|\}$, which produces $u^* = 2D\tilde{\gamma}/\gamma\,\nabla\log Z_\alpha = -\text{sign}(x)(2q\tilde{\gamma})^{1/2}/\gamma$, whose amplitude is independent of the coordinate $x$.

In this approximation one can calculate the maximum tolerance on the control amplitude $U$, given an allowed level of suboptimality. Using the results from the previous subsection one obtains

$$\frac{\delta \mathcal{F}_\alpha}{\mathcal{F}_\alpha^*} = \frac{1}{2(1 - 2D\alpha\gamma)}\left(\frac{\delta U}{U^*}\right)^2 = \frac{\chi}{4D}\left(\frac{\delta U}{U^*}\right)^2 \tag{34}$$

(see also Fig. 2). In the risk-neutral case, a relative error in the choice of $U^*$ of 10% results in a small increase of 0.5% for the total cost incurred. Risk-averse strategies tend to be less tolerant to errors, while risk-sensitive ones are more forgiving. We also remark that for $\alpha > 0$, the control amplitude is bound to be larger than a minimum value $U_{\min} = \sqrt{2D\alpha\gamma}\,U^*$, below which the risk-sensitive cost diverges.

### 1.5. Other forms for the cost of the control

The assumption that the cost for control is quadratic is particularly useful for two reasons. First, as we already remarked, it has a direct interpretation in terms of entropy and distance (Kullback–Leibler divergence) between ensembles of paths. Second, it has the property that the optimal control problem is linearly solvable (through the Hopf–Cole transformation the optimality equations can be cast into a linear form). Moreover, it has a physical interpretation, as the power dissipated moving in a viscous medium. Obviously, this is not the most general form for the control cost which can be physically motivated. One can use a generic function of $|u|$. For instance, the control cost of the form $\eta\,|u|$ corresponds, in the low noise limit, to the minimization of the path length to the target.

Here we derive the optimal HJB equation for a target location problem with a single agent and in the risk neutral case, where the control cost has the form an extra contribution which is linear in the control amplitude, $\eta\,|u|$. The minimization of the cost function constrained to the dynamics given by Eq. (1) is translated in the unconstrained minimization of the Lagrange functional

$$\mathcal{L}[u, p, \phi] = \int_0^\infty dt \int dx \left(\underbrace{\frac{\gamma}{2}u(x,t)^2 + \eta\,|u(x,t)|}_{\text{control}} + \underbrace{q(x)}_{\text{time}}\right)p(x,t) + \int_0^\infty dt \int dx\, \phi(x,t)\Big(\partial_t p + \nabla \cdot \big(u\,p\big) - D\nabla^2 p\Big)$$

The stationarity with respect to $u$ yields the equation for the optimal control

$$\gamma\,u^* + \eta\,\frac{u^*}{|u^*|} = \nabla\phi \tag{35}$$

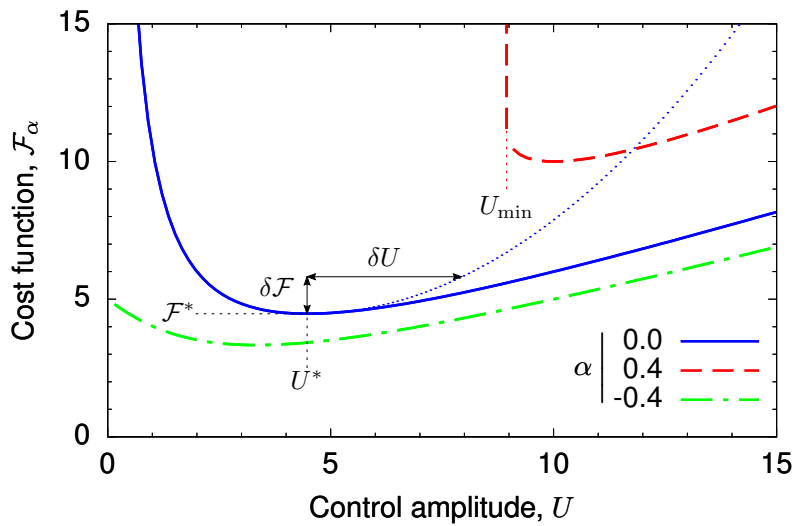

FIG. 2. **Robustness of the optimal solution.** The risk-sensitive cost function $\mathcal{F}_\alpha$ is plotted as a function of the control amplitude $U$, in a risk-neutral (solid blue), risk-averse (dashed red) and risk-seeking (dashed-dotted green) situation. The parabolic approximation around the minimum is shown for the risk-neutral case (dotted blue line). A deviation of the control from the optimum by a quantity $\delta U$ corresponds to an increase in the cost $\delta \mathcal{F}$ (in the parabolic approximation). These two quantities are related by Eq. (34). Curves have been obtained with the same parameters as in Figs. 2 and 4 of the main text.

and with respect to $p$ gives the optimality HJB equation

$$\partial_t \phi + D\nabla^2 \phi + \frac{1}{2\gamma}|\nabla\phi|^2 - \eta|\nabla\phi| = q - \frac{\eta^2}{2\gamma} \ . \tag{36}$$

Through the Hopf–Cole transformation $\phi = 2D\gamma \log Z$, the HJB equation becomes

$$\partial_t Z + D\nabla^2 Z + \eta|\nabla Z| = \frac{1}{2D\gamma}\left(q - \frac{\eta^2}{2\gamma}\right)Z \ . \tag{37}$$

The dynamics of the desirability (chemoattractant) $Z$ acquires a ballistic contribution, such that in addition to the diffusive motion it also propagates as a front.

## 2. MEAN-FIELD APPROXIMATION

The Bellman equation (26) is equivalent to the stationary Schrödinger equation with zero energy for a system of $N$ identical interacting particles. It seems impossible to solve it exactly with the boundary conditions discussed above. We therefore proposed a mean-field approximation scheme which is motivated both physically (for large number of particles and reasonably diluted systems) and from the game-theoretical point of view (inspired by mean-field games).

### 2.1. Risk-neutral case

For $\alpha = 0$ the mean-field approximation to this equation consists in factorizing the $N$-point solution $Z$ into the product of identical functions of the individual variables:

$$Z(x_1 \ldots x_N) \stackrel{\mathrm{MF}}{=} \prod_{i=1}^{N} \zeta(x_i) \ ; \tag{38}$$

from this mean-field ansatz it follows that the controls are

$$u_i^* = 2D\nabla \log \zeta(x_i) \ , \tag{39}$$

and that the $N$-particle density function is also factorized in single-particle distribution functions,

$$p(x_1 \ldots x_N) = \prod_{i=1}^{N} \rho(x_i) \tag{40}$$

(provided that the initial positions of the $N$ particles are independent); the single-particle density $\rho$ then satisfies

$$\partial_t \rho + \nabla \cdot (u\,\rho) = D\nabla^2 \rho \ . \tag{41}$$

One can then replace the ansatz in Eqs. (39) and (40) in the cost functional $\mathcal{F}$ to obtain the cost-per-particle functional

$$\tilde{\mathcal{F}} = \int dt \int dx \left(\frac{\gamma}{2}\,u(x,t)^2 + q(x) + \frac{N-1}{2}\int dy\,v(x,y)\rho(y,t)\right)\rho(x,t) \tag{42}$$

The mean-field optimal control equations follow by applying the Pontryagin principle to the functional $\tilde{\mathcal{F}}$ under the constraint (41), i.e. as the saddle point equations for the Lagrange functional

$$\tilde{\mathcal{L}} = \tilde{\mathcal{F}} + \int dt \int dx\,\phi(x,t)\left(\partial_t \rho + \nabla \cdot (u\,\rho) - D\nabla^2 \rho\right) \ , \tag{43}$$

where variations have to be calculated with respect to the single particle functions $u$, $\rho$ (and $\phi$, yielding the constraint). This leads to

$$\left.\frac{\delta \tilde{\mathcal{L}}}{\delta u(x,t)}\right|_* = \rho\left(\gamma u^* - \nabla\phi\right) = 0 \qquad \Rightarrow \qquad u^*(x,t) = \gamma^{-1}\nabla\phi(x,t) \ , \tag{44}$$

and

$$\left.\frac{\delta\tilde{\mathcal{L}}}{\delta\rho(x,t)}\right|_* = \frac{\gamma}{2}u^{*2} + h_{\mathrm{MF}} - \partial_t\phi - u^*\cdot\nabla\phi - D\nabla^2\phi$$

$$= -\partial_t\phi - \frac{1}{2\gamma}(\nabla\phi)^2 - D\nabla^2\phi + h_{\mathrm{MF}} = 0 \ , \tag{45}$$

where $h_{\mathrm{MF}}$ is the mean-field cost

$$h_{\mathrm{MF}}(x_i,t) = q(x_i) + (N-1)\int dy\, v(x_i,y)\,\rho(y,t) \ . \tag{46}$$

When $v$ is a contact interaction potential, $v(x,y) = g\,\delta(x-y)$, one has

$$h_{\mathrm{MF}} = q + g\,(N-1)\,\rho \ . \tag{47}$$

By applying the Hopf–Cole transformation $\phi = 2D\gamma\log\zeta$, one finally gets the HJB equation for the mean-field desirability, $\zeta$,

$$\partial_t\zeta - D\nabla^2\zeta = \frac{1}{2D\gamma}h_{\mathrm{MF}}\,\zeta \ , \tag{48}$$

which for contact potential reads

$$\partial_t\zeta - D\nabla^2\zeta = \frac{1}{2D\gamma}\Big(q + g(N-1)\rho\Big)\,\zeta \ . \tag{49}$$

### 2.1.1. Optimality of the cooperative solution

In this section we show that the best mean-field solution is indeed the cooperative one, i.e. the one in which the individual controls are identical.

Let us assume that out of the total $N$ agents, $N_1$ are of one species and $N_2 = N - N_1$ are of a second species, with different control. We can, for the sake of generality, introduce different collision costs depending on the species of the agents involved. Therefore, the cost rate for an agent of species $\alpha$ colliding with an agent of species $\beta$ is $g_{\alpha\beta}/2$. The mean-field costs incurred by individual agents of species 1 and 2 are

$$C_1 = \bar{C}_1 + \frac{1}{2}\int dt\,dt\Big[g_{11}(N_1-1)\,\rho_1^2 + g_{12}(N-N_1)\rho_1\,\rho_2\Big] \ , \tag{50a}$$

$$C_2 = \bar{C}_2 + \frac{1}{2}\int dt\,dt\Big[g_{21}\,N_1\,\rho_1\,\rho_2 + g_{22}(N-N_1-1)\rho_2^2\Big] \ , \tag{50b}$$

where $\bar{C}_\alpha$ is the control and time cost for an agent of species $\alpha$,

$$\bar{C}_\alpha = \int dt\,dx\,\rho_\alpha\left(\frac{\gamma_\alpha}{2}u_\alpha^2 + q_\alpha\right) \ . \tag{51}$$

The goal of each agent of species $\alpha$ is to maximize the cost $C_\alpha$. We shall see that if the collision costs do not depend on the species involved, i.e. $g_{ij} = g$, the best partition of the system is $N_1 = 0$ or $N_1 = N$.

One observes that $C_1$ and $C_2$ both have linear dependence on $N_1$:

$$\frac{\partial C_1}{\partial N_1} = \frac{g}{2}\int dt\,dt\Big[\rho_1^2 - \rho_1\,\rho_2\Big] \ , \tag{52a}$$

$$\frac{\partial C_2}{\partial N_1} = \frac{g}{2}\int dt\,dt\Big[\rho_1\,\rho_2 - \rho_2^2\Big] \ . \tag{52b}$$

If $C_1$ decreases with $N_1$ and $C_2$ increases with $N_1$ (i.e. decreases with $N_2$), or viceversa, then the two species should coexist. Assuming $\partial C_1/\partial N_1 < 0$ and $\partial C_2/\partial N_1 > 0$ would imply

$$0 > \frac{\partial C_1}{\partial N_1} - \frac{\partial C_2}{\partial N_1} = \frac{g}{2}\int dt\,dx\left[\rho_1 - \rho_2\right]^2$$

which is not possible. The same conclusion holds for $\partial C_1/\partial N_1 > 0$ and $\partial C_2/\partial N_1 < 0$. Therefore, one must have both $C_1$ and $C_2$ decreasing (or increasing) functions of $N_1$, which makes it more desirable to have $N_1 \to N$ (or $N_1 \to 0$) for both species.

### 2.1.2. Effect of the collision cost: travelling wave solution of PKS in 1D

As remarked in the main text, the optimal control equations in the mean-field approximation are equivalent to the Patlak–Keller–Segel (PKS) equations with logaritmic response. In this section we show that in the case where $q = 0$ (no time expenditure cost) it is possible to find travelling wave solution to the PKS equations in one dimension [7]. We shall see that the combination $g(N-1)$ enters the definition of the travelling wave velocity.

In one dimension, the optimal control equations for $q = 0$ are

$$\begin{cases} D\partial_x^2\zeta - \dfrac{1}{2D\gamma}\big(q + g(N-1)\rho\big)\zeta = 0 \;, \\ \partial_t\rho + 2D\partial_x\big(\rho\,\partial_x\zeta/\zeta\big) - D\partial_x^2\rho = 0 \;. \end{cases} \tag{53}$$

We impose the boundary conditions $\zeta(-\infty) = 0$, $\zeta(+\infty) = 1$, $\rho(-\infty) = \rho_\infty$ and $\rho(+\infty) = 0$. Such boundary conditions correspond to the situation in which a constant supply of agents is provided very far on the left and the target is far on the right. The system admit a travelling wave solution, $\zeta(x,t) = Z(x - ct)$ and $\rho(x,t) = R(x - ct)$, with velocity $c > 0$. With this ansatz, Eqs. (53) write

$$\begin{cases} DZ'' - \dfrac{g(N-1)}{2D\gamma}RZ = 0 \;, \\ -cR' + 2D\big(R\,Z'/Z\big)' - DR'' = 0 \;, \end{cases} \tag{54}$$

where the symbol $'$ indicates the derivative with respect to the single variable $z = x - ct$. The second equation in the system (54) can be straightforwardly integrated once and gives

$$R' = (2Z'/Z - \kappa)\,R + \beta \;,$$

where $\kappa = c/D$. If we impose the boundary conditions $R|_\infty = 0$ and $R'|_\infty = 0$, the integration constant $\beta$ vanishes. One further integration gives

$$\log R = 2\log Z - \kappa z + \alpha \qquad \Rightarrow \qquad R(z) = A\,e^{-\kappa z}\,Z(z)^2 \;.$$

By replacing this solution into the first equation of the system (54), and by defining $Z(z) = e^{\kappa z/2}\chi(z)$

$$D\chi'' + c\chi' + \dfrac{1}{2D}\Big(\dfrac{c^2}{2} - \dfrac{g(N-1)}{\gamma}\,A\chi^2\Big)\chi = 0 \;, \tag{55}$$

Notice that $R = A\,\chi^2$, and therefore the boundary conditions for $\chi$ follow from the ones for $\rho$. From the stability condition $c^2/2 - g(N-1)A\chi^2(-\infty)/\gamma = 0$, it follows that the speed of the wave front is

$$c = \sqrt{\dfrac{2\rho_\infty}{\gamma}\,g(N-1)} \;. \tag{56}$$

The number of agents therefore influences the speed at which the agent density profile propagates: more agents consume the chemoattractant more rapidly, hence giving rise to steepest gradients of its concentration $\zeta$, which then yields stronger drift.

## 2.2. Risk-sensitive case

We now derive the mean-field equation for the risk-sensitive case, in particular the risk-averse one, $\alpha > 0$ (easily extended to the risk-seeking one, $\alpha < 0$). In this case, the mean-field ansatz consists in assuming that the single-particle contributions to the total cost are independent and identically distributed. We start by writing the evolution of the joint stochastic process of particle positions and individual costs as the $2N$ coupled equations

$$\begin{cases} dX_i^t = u_i\big(X_1^t, \dots X_N^t\big)\,dt + \sqrt{2D}dW_i^t \;, \\ dC_i^t = \Big(\dfrac{\gamma}{2}u_i(X_1^t, \dots X_N^t, t)^2 + q(X_i^t) + \dfrac{1}{2}\displaystyle\sum_{j \neq i} v(X_i^t, X_i^t)\Big)\,dt \equiv c_i\big(X_1^t, \dots X_N^t, t\big)\,dt \;, \end{cases} \tag{57}$$

The $N$-particle position-cost density $p_{\text{XC}}(x_1, C_1 \ldots x_N, C_N, t)$ associated to Eqs. (57) is also factorized

$$p_{\text{XC}}(x_1, C_1 \ldots x_N, C_N, t) \overset{\text{MF}}{=} \prod_i \rho_{\text{XC}}(x_i, C_i, t) \ , \tag{58}$$

following from the assumption that the controls are given by a unique function of the individual one-particle positions, namely

$$u_i(X_1^t, \ldots X_N^t, t) \overset{\text{MF}}{=} u(X_i^t, t) \ . \tag{59}$$

It follows that the cost functional $\mathcal{G}_\alpha$ also factorizes:

$$\mathcal{G}_\alpha = \int dx_1 \, dC_1 \ldots dx_N \, dC_N \, p_{\text{XC}}(x_1, C_1 \ldots x_N, C_N, T) \, e^{\alpha \sum_i C_i} \overset{\text{MF}}{=} \left( \int dx \, dC \, \rho_{\text{XC}}(x, C, T) e^{\alpha C} \right)^N \ . \tag{60}$$

The Fokker–Planck equation associated with Eqs. (57) is

$$\partial_t p_{\text{XC}} + \sum_i \partial_{C_i} \big( c_i(\bar{x}, t) \, p_{\text{XC}} \big) + \sum_i \nabla_i \big( u_i \, p_{\text{XC}} - D \nabla_i p_{\text{XC}} \big) = 0 \ , \tag{61}$$

and can be marginalized to the single-particle one by integrating over all particles but one:

$$\partial_t \rho_{\text{XC}} + \nabla \cdot (u \rho_{\text{XC}}) + \partial_C \left[ \left( \frac{\gamma}{2} u^2 + q + \frac{N-1}{2} \int dx' \, dC' \, v(x, x') \rho_{\text{XC}}(x', C') \right) \rho_{\text{XC}} \right] - D \nabla^2 \rho_{\text{XC}} = 0 \ . \tag{62}$$

The mean-field optimal control equations are then derived (applying Pontryagin principle) as the saddle point equations of the functional

$$\tilde{\mathcal{L}}_\alpha = \int dx \, dC \, \rho_{\text{XC}}(x, C, T) e^{\alpha C} + \int dx \, dC \, dt \, \chi_\alpha(x, C, t) \Big\{ \partial_t \rho_{\text{XC}}(x, C, t) + \nabla \cdot (u(x, t) \rho_{\text{XC}}(x, C, t)) - D \nabla^2 \rho_{\text{XC}}(x, C, t)$$
$$+ \partial_C \left[ \left( \frac{\gamma}{2} u(x, t)^2 + q(x) + \frac{N-1}{2} \int dx' \, dC' \, v(x, x') \rho_{\text{XC}}(x', C', t) \right) \rho_{\text{XC}}(x, C, t) \right] \Big\} \tag{63}$$

Variation with respect to the control yields

$$\frac{\delta \tilde{\mathcal{L}}_\alpha}{\delta u(x, t)} \bigg|_* = -\int dC \, \rho_{\text{XC}}(x, C, t) \Big[ \nabla \chi_\alpha(x, C, t) + \gamma u(x, t) \partial_C \chi_\alpha(x, C, t) \Big] = 0 \ , \tag{64}$$

and with respect to $\rho_{\text{XC}}$

$$\frac{\delta \tilde{\mathcal{L}}_\alpha}{\delta \rho_{\text{XC}}(x, C, t)} \bigg|_* = e^{\alpha C} \, \delta(t - T) - \Big( \partial_t \chi_\alpha + u^* \cdot \nabla \chi_\alpha + D \nabla^2 \chi_\alpha + (\gamma u^{*2}/2 + h_{\text{MF}}) \partial_C \chi_\alpha \Big) = 0 \ , \tag{65}$$

which is the (backward) equation for the functional $\chi_\alpha(x, C, t) = \big\langle \exp \alpha C_0^T \big| X^t = x, \, C_0^t = C \big\rangle_*$, where $h_{\text{MF}}$ is the mean-field cost Eq. (46).

$$h_{\text{MF}}(x, t) = q(x) + (N-1) \int dx' \underbrace{\int dC' \rho_{\text{XC}}(x', C', t)}_{\rho(x', t)} v(x, x') \ . \tag{66}$$

The $\delta$-function at the final time sets the condition $\chi_\alpha|_{t=T} = e^{\alpha C}$. We notice that the function $\chi_\alpha$ can be expressed as

$$\chi_\alpha(x, C, t) = \big\langle e^{\alpha(C_0^t + C_t^T)} \big| X^t = x, \, C_0^t = C \big\rangle_* = e^{\alpha C} \big\langle e^{\alpha C_t^T} \big| X^t = x \big\rangle_* \equiv e^{\alpha C} \psi_\alpha(x, t) \tag{67}$$

We therefore see that the optimal control can be written in terms of $\psi_\alpha(x, t)$

$$u^*(x, t) = \frac{1}{\gamma} \nabla \left( -\frac{1}{\alpha} \log \psi_\alpha(x, t) \right) \ , \tag{68}$$

i.e. as the gradient of the risk-sensitive (mean-field) value function $\phi_\alpha = -\alpha^{-1} \log \psi_\alpha$, which satisfies the HJB equation

$$\partial_t \phi_\alpha + \frac{1}{2\tilde{\gamma}} \big( \nabla \phi_\alpha \big)^2 + D \nabla^2 \phi_\alpha = h_{\text{MF}} \ , \tag{69}$$

where we recall from the previous section that $\tilde{\gamma} = \gamma/(1 - 2D\alpha\gamma)$. The mean-field desirability $\zeta_\alpha = \exp(\phi_\alpha/2D\tilde{\gamma})$ then satisfies the linear HJB equation

$$\partial_t \zeta_\alpha + D \nabla^2 \zeta_\alpha = \frac{1}{2D\tilde{\gamma}} h_{\text{MF}} \, \zeta_\alpha \ . \tag{70}$$

## 3. EXACT SOLUTION FOR THE CIRCULAR TARGET IN THE NON-INTERACTING CASE

It is possible to solve analytically the HJB equation (for the desirability, in the terminal state setup) for the search problem of a circular target in the infinite two-dimensional space, in absence of interactions. In this case, the mean-field ansatz is trivially exact, provided that the particles are independently distributed at the initial time. If the target has radius $R$ and we choose the origin of the coordinate system to be its center, the HJB equation in cylindrical coordinates reads

$$\frac{D}{r}\partial_r\left(r\partial_r\zeta_\alpha\right) - \frac{q}{2D\tilde{\gamma}}\,\zeta_\alpha = 0 \ , \tag{71}$$

where the desirability $\zeta_\alpha$ depends only on the radial coordinate, and, from previous sections, $\tilde{\gamma} = \gamma/(1 - 2D\alpha\gamma)$.

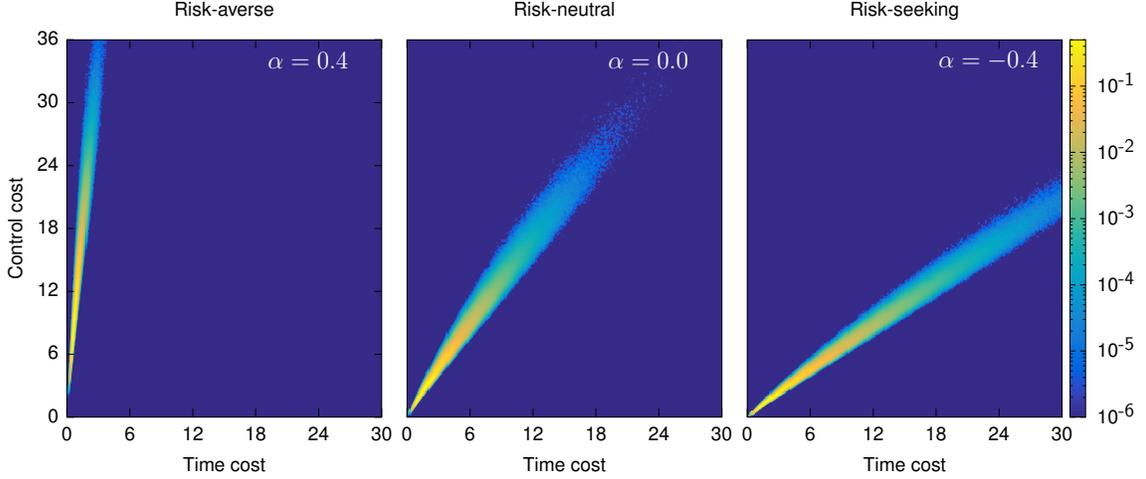

FIG. 3. **Risk-neutral vs Risk-sensitive.** Two-dimensional histograms of time-expenditure and control costs, from simulations of Eq. (1) with the mean-field control in Eq. (73), for the risk-averse, risk-neutral and risk-seeking situation. The costs for time-expenditure and control are positively and almost linearly correlated. The optimal control for the risk-neutral problem is such that control and time-expenditure costs are very similar. Instead, the risk-averse optimal controller tends to pay more on control (reducing possible dangerous fluctuations towards high values of the cost), whereas the risk-seeking controller allows for large time-expenditure cost while reducing the control.

Given the connection of the desirability with the expected cost-to-go function (see main text), the boundary conditions for Eq. (71) are $\zeta(r \to \infty) = 0$ and $\zeta(R) = 1$. The solution to this problem is

$$\zeta_\alpha(r) = \frac{K_0\left(\sqrt{\frac{q}{2D^2\tilde{\gamma}}}\,r\right)}{K_0\left(\sqrt{\frac{q}{2D^2\tilde{\gamma}}}\,R\right)} \ , \tag{72}$$

which yields the optimal control

$$u = \frac{2D\tilde{\gamma}}{\gamma}\nabla\log\zeta_\alpha = -\frac{\sqrt{2\tilde{\gamma}q}}{\gamma}\frac{K_1\left(\sqrt{\frac{q}{2D^2\tilde{\gamma}}}\,r\right)}{K_0\left(\sqrt{\frac{q}{2D^2\tilde{\gamma}}}\,r\right)}\,\hat{e}_r \ , \tag{73}$$

where $K_\nu$ are the modified Bessel functions of second kind and $\hat{e}_r \equiv x/r$, i.e. the outward unit vector pointing to $x$ from the origin.

In Fig. 3, one can see that the risk-sensitivity parameter sets an imbalance between the control cost and the time-expenditure cost: risk-averse strategies are prone to pay much more price on control than on time-expenditure, while for risk-neutral strategies the difference is much less pronounced; risk-seeking controllers, instead, pay less cost in

control, confident of being driven to the target by the noise.

---



\end{document}